\documentclass[preprint2]{aastex}

\usepackage{amsmath}
\usepackage[colorlinks=true,allcolors=blue,urlcolor=magenta,bookmarks=false,pdfstartview=FitV]{hyperref}

\shorttitle{Dust diffusion and settling in the presence of collisions}
\shortauthors{Krijt \& Ciesla}

\begin{document}

\title{Dust diffusion and settling in the presence of collisions: \\Trapping (sub)micron grains in the midplane*}

\author{Sebastiaan Krijt and Fred J. Ciesla}
\email{skrijt@uchicago.edu}
\affil{Department of the Geophysical Sciences, The University of Chicago, 5734 South Ellis Avenue, Chicago, IL 60637, USA}

\altaffiltext{*}{Accepted for publication in ApJ}

\begin{abstract}
In protoplanetary disks, the distribution and abundance of small (sub)micron grains are important for a range of physical and chemical processes. For example, they dominate the optical depth at short wavelengths and their surfaces are the sites of many important chemical reactions such as the formation of water. Based on their aerodynamical properties (i.e., their strong dynamical coupling with the surrounding gas) it is often assumed that these small grains are well-mixed with the gas. Our goal is to study the vertical (re)distribution of grains taking into account settling, turbulent diffusion, as well as collisions with other dust grains. Assuming a fragmentation-limited background dust population, we developed a Monte Carlo approach that follows single monomers as they move through a vertical column of gas and become incorporated in different aggregates as they undergo sticking and fragmenting collisions. We find that (sub)micron grains are not necessarily well-mixed vertically, but can become trapped in a thin layer with a scale-height that is significantly smaller than that of the gas. This collisional trapping occurs when the timescale for diffusion is comparable to or longer than the collision timescale in the midplane and its effect is strongest when the most massive particles in the size-distribution show significant settling. Based on simulations and analytical considerations we conclude that for typical dust-to-gas ratios and turbulence levels, the collisional trapping of small grains should be a relatively common phenomenon. The absence of trapping could then indicate a low dust-to-gas ratio, possibly because a large portion of the dust mass has been removed through radial drift or is locked up in planetesimals.
\end{abstract}


\section{Introduction}
Initially, protoplanetary disks are believed to consist of gas and microscopic dust particles, mixed at a mass ratio of about 100:1 \citep{williamscieza2011}, resembling the interstellar medium. The relatively high densities inside the protoplanetary disk make dust coagulation on timescales (much) shorter than $10^{6-7}\mathrm{~yr}$ (the typical lifetime of the protoplanetary disk) possible and the growth of microscopic particles into centimeter-size pebbles and meter-size boulders constitute the first steps of planet formation in the standard core accretion scenario \citep{testi2014,johansen2014}. While the dust size distributions that arise from coagulation can have complex shapes and vary greatly with time and location \citep[e.g.,][]{dullemonddominik2005,birnstiel2011}, it is usually true that the total mass is dominated by the largest particles while small dust grains provide the majority of the surface area for solids. And while the larger, mass-dominating particles are interesting as potential building blocks of planetesimals and eventually planets, the small (sub)micron grains are important because they dominate the optical depth at UV wavelengths, influencing the temperate structure of the gas \citep{dutrey2014} and their surface plays an important role in many chemical processes including the formation of water molecules \citep[see][for recent reviews]{pontoppidan2014,vandishoeck2014}.

In the inner disk, where collision times are short and collision velocities are high, the current picture is that small dust grains are continuously created in destructive collisions of larger, mm to cm-size bodies, resulting in a steady-state between fragmentation and coagulation. Without these destructive collisions, the abundance of small grains will drop very rapidly after coagulation kicks in \citep{dullemonddominik2005}. Even though small grains are predominantly created close to the midplane, where the larger bodies are concentrated, it is generally held that they will be efficiently mixed vertically. Specifically, their vertical scale-height is determined by balancing settling and vertical diffusion \citep{cuzzi1993,dubrulle1995,youdin2007,ciesla2010}. Since small grains couple to the gas fairly well, settling is not very effective and their scale height is usually taken to be equal to the gas scale height.

We set out to model the interaction of diffusion, settling, and collisional processes, and understand how individual grains move vertically under realistic conditions, focusing on those regions of the disk where the dust size distribution is expected to be in a growth/fragmentation equilibrium. In particular, we want to see what effect particle growth has on how a single monomer is transported vertically in the disk, be it alone or as part of a larger aggregate. The paper is structured as follows. Section \ref{sec:disk_dust} outlines the models used to describe the protoplanetary nebula and steady-state dust distribution. In Sect. \ref{sec:method} we extend the purely dynamical model of \citet{ciesla2010} to include particle-particle collisions. In Sect. \ref{sec:individual} individual dust grains are followed and Sect. \ref{sec:collective} combines multiple grain histories to quantify the vertical distribution of small grains. The results and their implications are discussed in Sect. \ref{sec:disc} and conclusions summarized in Sect. \ref{sec:concl}.

\section{Disk model and background dust population}\label{sec:disk_dust}
We must first define models for the protoplanetary disk and the background dust population. For now, the properties of the nebula and the global shape of the dust population it houses are assumed to be time-independent, though the approach developed here can be used with time-dependent models as well.

\subsection{Disk model}
Focusing on a young protoplanetary disk around a Sun-like star, the radial gas surface density and temperature profiles are taken to equal
\begin{equation}\label{eq:Sigma_g}
\Sigma_\mathrm{g} = 2000\mathrm{~g~cm^{-2}} \left( \frac{r}{1\mathrm{~AU}} \right)^{-3/2},
\end{equation}
and
\begin{equation}\label{eq:T}
T = 280\mathrm{~K} \left( \frac{r}{1\mathrm{~AU}} \right)^{-1/2},
\end{equation}
consistent with the Minimum Mass Solar Nebula \citep{hayashi1981}. The gas diffusivity is taken to equal
\begin{equation}
D_\mathrm{g} = \alpha c_s h_\mathrm{g},
\end{equation}
where the gas scale height is given by $h_\mathrm{g} = c_s / \Omega$ with $\Omega$ the Keplerian frequency, $c_s=(k_\mathrm{B}T/m_\mathrm{g})^{1/2}$ the isothermal sound speed, and $m_\mathrm{g}=3.9\cdot10^{-24}\mathrm{~g}$ denotes the mean molecular weight. Following \citet{shakura1973}, we use the dimensionless parameter $\alpha$ to describe the strength of the turbulence. 

In the isothermal case, $D_\mathrm{g}$ does not depend on height and the gas density drops away from the midplane according to
\begin{equation}\label{eq:rho_g}
\rho_\mathrm{g} = \frac{\Sigma_\mathrm{g}}{\sqrt{2\pi} h_\mathrm{g}} \exp\left\{ -\frac{z^2}{2 h_\mathrm{g}^2} \right\},
\end{equation}
 The timescale on which turbulent diffusion occurs equals
\begin{equation}\label{eq:t_mix}
t_D = \frac{h^2_\mathrm{g}}{D_\mathrm{g}} = \frac{1}{\alpha \Omega} \propto r^{3/2},
\end{equation}
and is longest for low turbulence strengths. 

\subsection{Dust particle properties}
The smallest grains considered are monomers with a radius $s_\bullet = 0.1\mathrm{~\mu m}$ and an internal density of $\rho_\bullet = 1.4\mathrm{~g~cm^{-3}}$. Larger particles\footnote{A brief note on the nomenclature used in this work: \emph{Monomers} refers specifically to these $0.1\mathrm{~\mu m}$ particles. \emph{Aggregates} are conglomerates of monomers while \emph{grains} usually refers to small particles, which could be monomers or small aggregates.} (aggregates of many monomers held together by surface forces) are assumed to be compact and spherical in shape, so that their mass and radius are related through $m = (4/3)\pi s^3 \rho_\bullet$. The size of the largest aggregates is set by the fragmentation threshold velocity and the properties of the disk (Sect. \ref{sec:background}).

A dust particle's dynamics are governed by its stopping time
\begin{equation}\label{eq:t_s}
t_s = \sqrt{\frac{\pi}{8}} \frac{\rho_\bullet s}{\rho_\mathrm{g} c_s},
\end{equation}
assuming we are in the Epstein limit, i.e., $s < (9/4) \lambda_\mathrm{mfp}$, with $\lambda_\mathrm{mfp} = m_\mathrm{g} / (\sigma_\mathrm{mol} \rho_\mathrm{g})$ the gas molecule mean free path. Because of the higher gas density, $\lambda_\mathrm{mfp}$ is lowest in the midplane where $\lambda_\mathrm{mfp} \sim 1\mathrm{~cm}$ at $r=5\mathrm{~AU}$ for the disk model considered here. The dimensionless form of the the stopping time is often defined as the Stokes number $\mathrm{St}\equiv \Omega t_s$.

The dust particle diffusivity is related to $D_\mathrm{g}$ through the Schmidt number \citep{youdin2007}
\begin{equation}
\mathrm{Sc} \equiv \frac{D_\mathrm{g}}{D_\mathrm{d}} \sim 1 + (\Omega t_s)^2.
\end{equation}
In addition to diffusion, dust grains will settle to the midplane of the disk by balancing the force of gravity with the resisting drag force of the gas, yielding a velocity of
\begin{equation}
v_\mathrm{settle} = -t_s \Omega^2 z.
\end{equation}
When diffusion and settling are balanced, the density distribution of dust particles of a given size will describe a Gaussian (equivalent to Eq. \ref{eq:rho_g}) with a relative scale height \citep{youdin2007}
\begin{equation}\label{eq:h_d}
\frac{h_\mathrm{d}}{h_\mathrm{g}} \approx \sqrt{ \frac{\alpha}{\alpha+\Omega t_s} \left( \frac{1+\Omega t_s}{1+2\Omega t_s} \right) }
\end{equation}
Thus, well-coupled grains follow the gas distribution (e.g., $h_\mathrm{d} \sim h_\mathrm{g}$), while particles with Stokes numbers $\Omega t_s > \alpha$ will settle toward the midplane.

\begin{figure}[!t]
\centering
\includegraphics[clip=,width=1.\linewidth]{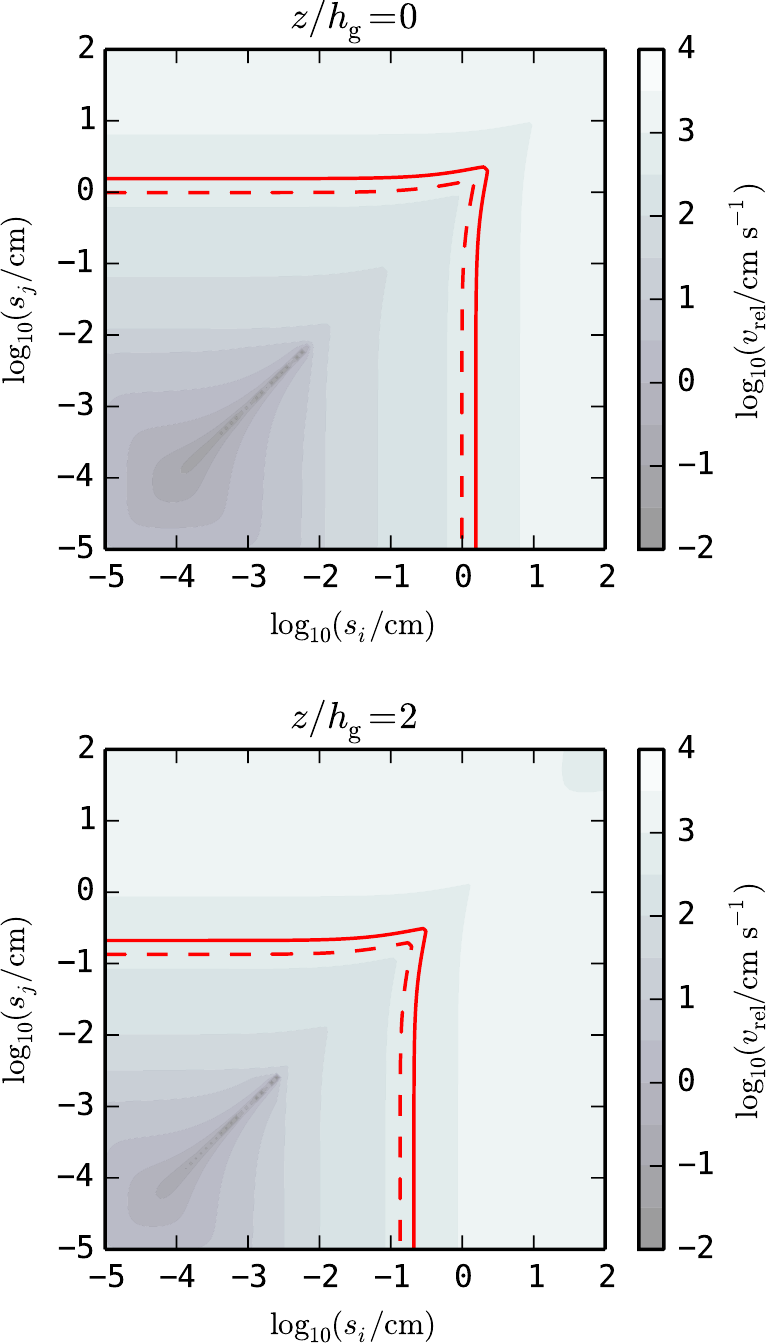}
\caption{Collision velocities at $5\mathrm{~AU}$ and $\alpha=10^{-3}$ at the midplane (top panel) and at $z/h_\mathrm{g}=2$ (bottom panel). Brownian motion, turbulence, and differential settling have been included. The red contours mark $v_\mathrm{frag}$ (solid) and $v_\mathrm{frag}-\delta v_\mathrm{frag}$ (dashed).}
\label{fig:vels}
\end{figure}

The collision velocity between two grains is calculated by adding contributions from Brownian motion and turbulence quadratically\footnote{For consistency with \citet{birnstiel2011} in Sect. \ref{sec:background}, the contribution of differential vertical settling is ignored. For the turbulence strengths considered in this work, this approximation is justified since the turbulent term will generally dominate over the differential settling velocity \citep[see for example][Fig. 2]{krijt2015}.}
\begin{equation}
v_\mathrm{rel} = \sqrt{ (\Delta v_\mathrm{BM})^2 +  (\Delta v_\mathrm{tur})^2 }.
\end{equation}
For $\Delta v_\mathrm{tur}$ we use Eq. 16 of \citet{ormel2007b} and 
\begin{equation}
\Delta v_\mathrm{BM} = \sqrt{  \frac{8 k_\mathrm{b} T (m_i+m_j)}{\pi m_i  m_j} },
\end{equation}
with $m_i$ and $m_j$ the masses of the two particles in question. Figure \ref{fig:vels} shows the relative collision velocity between two particles at $5\mathrm{~AU}$ assuming that this region of the disk can be described by the parameters given in Table \ref{tab:benchmark}. The upper panel shows the midplane collision velocity, and the lower panel shows $v_\mathrm{rel}$ at $z=2h_\mathrm{g}$. The main reason for the differences in relative velocities is the lower value of $\rho_\mathrm{g}$ away from the midplane, which -- for a fixed particle size -- leads to higher Stokes numbers.

\subsection{Sticking, fragmentation, and erosion}\label{sec:collisions}
Dust particles can gain and lose mass through collisions with other dust grains. In order for two aggregates to stick, they must be able to absorb and dissipate the collisional energy without falling apart. At low velocities, when the collision energy is small compared to the binding energy of the aggregates, this is generally not a problem. At high velocities however, when the kinetic energy becomes comparable or larger than the aggregate's binding energy, fragmentation becomes the dominant outcome. In the last 2 decades, numerous authors have studied the transition from sticking to fragmentation experimentally \citep[e.g.,][and references therein]{blumwurm2000,blumwurm2008,guttler2010} and through numerical simulations \citep[e.g.,][]{dominiktielens1997,ringl2012,wada2013}, focusing on the influence of collider sizes, shapes, and materials.

We employ the collision model of \citet{birnstiel2011} and assume the fragmentation probability for a collision is given by
\begin{equation}\label{eq:P_frag}
P_\mathrm{frag}=
\begin{cases}
0 & \textrm{if~}  v_\mathrm{rel} < v_\mathrm{frag} - \delta v_\mathrm{frag}, \vspace{3mm} \\ 
1 & \textrm{if~}  v_\mathrm{rel} \geq v_\mathrm{frag}, \vspace{3mm} \\
1-\dfrac{v_\mathrm{frag}-v_\mathrm{rel}}{\delta v_\mathrm{frag}} & \textrm{at intermediate $v_\mathrm{rel}$,}
\end{cases}
\end{equation}
where $\delta v_\mathrm{frag} = v_\mathrm{frag}/5$. The intermediate transition regime with width $\delta v_\mathrm{frag}$ was included by \citet{birnstiel2011} because experimental investigations did not reveal a sharp transition from sticking to fragmentation \citep[e.g.,][]{blum1993}. Figure \ref{fig:vels} illustrates the boundaries of the pure sticking and fragmentation regimes with the dashed and solid red lines, showing that these regimes are a function of height in the disk and that the intermediate regime (where \emph{both} sticking and fragmentation can occur) covers a relatively narrow range of sizes. 

When sticking occurs, the size of the new dust aggregate is found by adding the two colliding masses. For fragmentation, we define two regimes: \emph{i)} if the mass ratio of the colliders is $R_m \geq 0.1$, catastrophic fragmentation occurs, and the mass of both colliders is redistributed over fragments; and \emph{ii)} when the mass ratio is $R_m<0.1$, erosion occurs, where the smaller collider excavates a mass equal to its own mass from the larger body. In both cases, the fragment distribution follows a power-law shape
\begin{equation}
n_f(s) = \begin{cases}
~ C_f s^{-\xi}  &\textrm{~for~} s_\bullet \leq s \leq s_{f,\mathrm{max}}, \vspace{3mm} \\ 
~ 0 &\textrm{~else},
\end{cases}
\end{equation}
where the size of the largest fragment $s_{f,\mathrm{max}}$ equals the larger collider in the case of catastrophic fragmentation, and the smaller collider in the case of erosion. The constant $C_f$ is determined by fixing the total mass of fragments to twice the mass of the smaller collider (for erosion), or to the sum of both collider masses (for catastrophic fragmentation). Throughout this work, we adopt $\xi=3.5$ as used by \citet{birnstiel2011}. For this power-law index, the fragment mass is dominated by the largest fragments, and the surface area by the smaller fragments.

\subsection{Background dust population}\label{sec:background}
We assume a steady-state population of dust particles with sizes ranging from $s_\bullet$ to some maximum size $s_\mathrm{max}$. Assuming growth is limited by turbulence-induced fragmentation, the maximum aggregate size is approximately \citep{birnstiel2009}
\begin{equation}\label{eq:s_max}
s_\mathrm{max} \approx \frac{2 \Sigma_\mathrm{g}}{\pi \alpha \rho_\bullet} \left( \frac{v_\mathrm{frag}}{c_s} \right)^2,
\end{equation}
where we have assumed the dust particle density is equal to $\rho_\bullet$ (i.e., compact aggregates).

\begin{figure}[t]
\centering
\includegraphics[clip=,width=.95\linewidth]{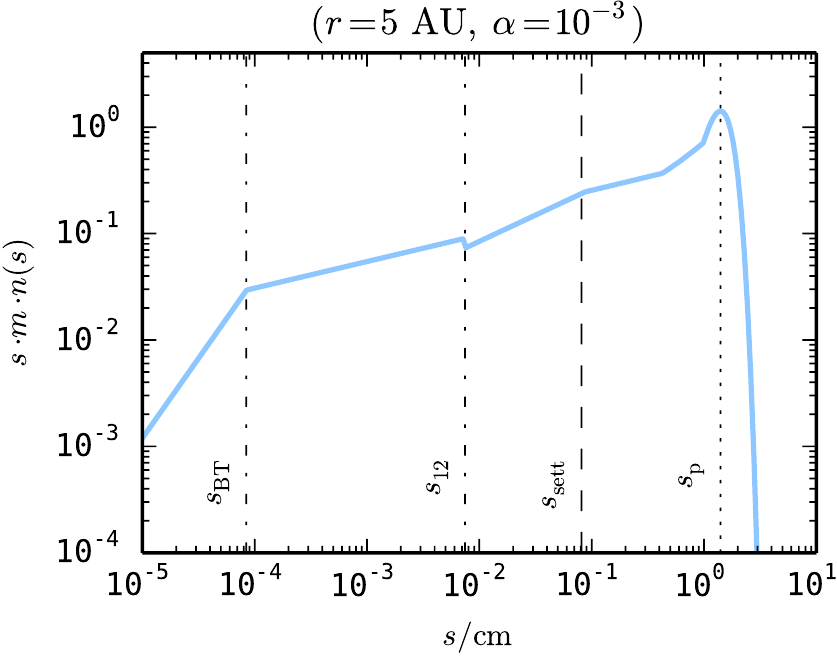}
\caption{Normalized steady-state dust size distribution derived for a fragmentation threshold of $v_\mathrm{frag}=5\mathrm{~m~s^{-1}}$ and the parameters of Table \ref{tab:benchmark}, using the approach of \citet{birnstiel2011} and as summarized in Sect. \ref{sec:background}.}
\label{fig:nama}
\end{figure}

\begin{table}
\begin{center}
\caption{The benchmark model used throughout this work.\label{tab:benchmark}}
\begin{tabular}{l c c}
\tableline \tableline
Quantity & symbol & value \\
\hline
Column location & $r$ & $5\mathrm{~AU}$\\
Gas surface density & $\Sigma_\mathrm{g}$ & $180\mathrm{~g~cm^{-2}}$ \\
Dust-to-gas ratio & $\Sigma_\mathrm{d}/\Sigma_\mathrm{g}$ & $0.01$ \\
Turbulence strength & $\alpha$ & $10^{-3}$ \\
Monomer radius & $s_\bullet$ & $0.1\mathrm{~\mu m}$ \\
Monomer density & $\rho_\bullet$ & $1.4\mathrm{~g~cm^{-3}}$\\
Fragmentation velocity & $v_\mathrm{frag}$ & $5\mathrm{~m~s^{-1}}$\\
Fragment power-law & $\xi$ & 3.5 \\
\tableline
\end{tabular}
\end{center}
\end{table}

We determine the complete dust size distribution by using the vertically integrated steady-state recipe of \citet[][Sect. 5]{birnstiel2011}, derived for the same collisional model as summarized in Sect. \ref{sec:collisions}. In essence, this method consists of identifying a number of important grain sizes, and building the full size distribution by using appropriate power-law distributions between these sizes. Figure \ref{fig:nama} shows the resulting size-distribution for the model of Table \ref{tab:benchmark}. Apart from the maximum grain size $s_\mathrm{max}$, important sizes are: the size where relative velocities start to be dominated by turbulence rather than Brownian motions ($s_\mathrm{BT}$); the size above which grains start to settle ($s_\mathrm{sett}$); the size for which the stopping time becomes larger than the turnover time of the smallest eddies ($s_{12}$); and the size where the mass-distribution peaks ($s_\mathrm{p}$). For a more in-depth discussion of these sizes and how they influence the size distribution the reader is referred to \citet{birnstiel2011}. The advantage of using a pre-determined background dust population instead of calculating it self-consistently \citep[e.g.,][]{zsom2011}, is that we have a smooth and well-characterized size distribution at all heights $z$, something that is difficult to achieve in numerical models.

\section{Method}\label{sec:method}
We now turn our attention to a single monomer, located somewhere in a column of gas at a radius $r$ from the central star. The goal is to follow the dynamical evolution of this monomer while it moves vertically and becomes part of different aggregates that experience sticking and destructive collisions. 

\subsection{Vertical motions}\label{sec:diffbehav}
Inside the column, the vertical motions will be governed by the aerodynamical properties of the aggregate in which the monomer is incorporated and can be calculated using the methodology described in \citet{ciesla2010}. Basically, the location of a grain after a time step $\Delta t$ can be obtained through

\begin{equation}\label{eq:z_new}
z(t+\Delta t) = z(t) + v_\mathrm{eff} \Delta t + \mathcal{R}_1 \left[ \frac{2}{\zeta} D_\mathrm{d} \Delta t \right]^{1/2},
\end{equation}
with $\zeta=1/3$ and $\mathcal{R}_1$ is a random number between $[-1,1]$. The effective velocity is given by
\begin{equation}
v_\mathrm{eff} = v_\mathrm{gas} + v_\mathrm{settle} =  - D_\mathrm{d} \frac{z}{h_\mathrm{g}^2}  - t_s \Omega^2 z,
\end{equation}
where the second term on the RHS depends on both particle size and ambient gas density through the stopping time $t_s$. Unlike the settling velocity, the $v_\mathrm{gas}$ term does not represent a physical velocity, but follows mathematically from expanding the diffusion equation while taking into account that the gas density drops away from the midplane \citep{ciesla2010}.

\begin{figure*}[!t]
\centering
\includegraphics[clip=,width=1.\linewidth]{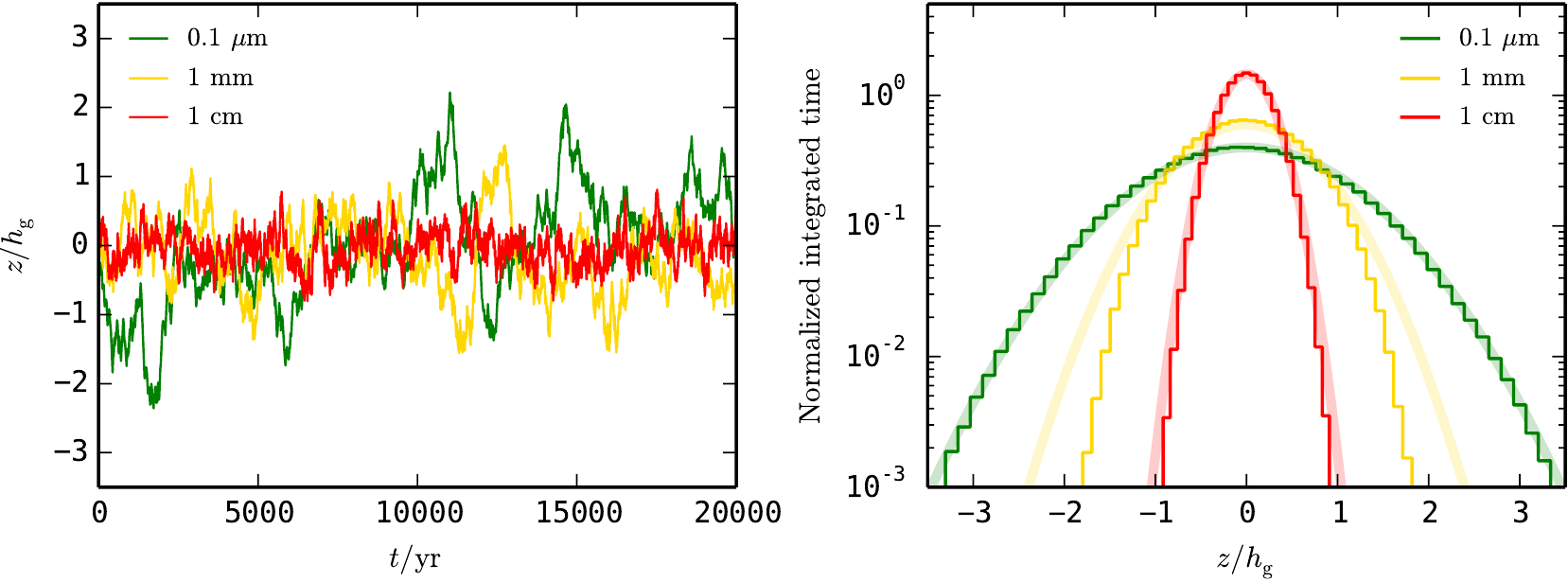}
\caption{Vertical motions of dust grains in the absence of collisions for particles with sizes $s=0.1\mathrm{~\mu m},~1\mathrm{~mm},~1\mathrm{~cm}$. Trajectories are simulated at $r=5\mathrm{~AU}$, assuming $\alpha=10^{-3}$. \emph{Left:} Three typical trajectories plotted as a function of time. \emph{Right:} Normalized histograms of the time spent at different heights. For each histogram, the histories of $10^3$ particles were combined. The shaded curves show predictions of Eq. \ref{eq:h_d} using the corresponding midplane Stokes numbers.}
\label{fig:noncol}
\end{figure*}

To ensure the time-step is not larger than a fraction of  the settling or diffusion timescale, we use
\begin{equation}
\Delta t = \frac{f_\mathrm{diff}}{(\alpha + \Omega t_s)\Omega},
\end{equation}
with $f_\mathrm{diff}=10^{-2}$.

To test the treatment of the vertical diffusion and settling, we simulate the motions of particles of three different sizes ($0.1\mathrm{~\mu m},~1\mathrm{~mm},~1\mathrm{~cm}$) in the absence of collisions. At $r=5\mathrm{~AU}$, these sizes have midplane Stokes numbers $\Omega t_s \sim 10^{-7},~10^{-3}$ and $10^{-2}$ respectively, and thus correspond to well-coupled ($\Omega t_s \ll \alpha$), marginally decoupled ($\Omega t_s \sim \alpha$), and settled ($\Omega t_s > \alpha$) particles. The left panel of Figure \ref{fig:noncol} shows the trajectories 3 different particles take in 20,000 yrs (approximately 10 mixing timescales $t_D$), and shows that, as expected, smaller particles make longer and more frequent excursions away from the midplane than larger particles. The right-hand panel shows normalized histograms of the time spent at different heights and compares them to predictions of Eq. \ref{eq:h_d}. For the histograms, histories of $10^3$ particles were combined for every size. The grains behave as one would expect, although, as noted by \citet{ciesla2010}, Eq. \ref{eq:h_d} slightly overestimates the width of the Gaussian because it neglects variations of the Stokes number with $z$, which our model does take into account.

\subsection{Collisional behavior}\label{sec:colbehav}
The next step is to include interactions between the aggregate that holds the monomer we are following and the background dust population of Sect. \ref{sec:background}. For this, the dust population is divided into $N_f$ size bins, with the characteristic sizes $s_j$ spaced logarithmically. The number density of grains of in size bin $j$ is given by
\begin{equation}\label{eq:n_j}
n_j = \frac{\rho_j}{m_j}  = \frac{1}{\sqrt{2\pi}h_{\mathrm{d},j} }   \frac{\Sigma_{\mathrm{d},j}}{m_j}  \exp \left\{ \frac{-z^2}{2 h_{\mathrm{d},j}^2 }  \right\},
\end{equation}
with $m_j=(4/3)\pi s_j^3 \rho_\bullet$. The surface densities, taken from the size-distribution of Sect. \ref{sec:background}, are normalized to add up to the total dust surface density
\begin{equation}
\sum_j^{N_f} \left( \Sigma_{\mathrm{d},j}\right) = \Sigma_\mathrm{d},
\end{equation}
which is related to $\Sigma_\mathrm{g}$ through the assumed (and time-independent) dust-to-gas ratio.

The collision rate between the (single) particle of size $s$ and a single particle in size bin $j$ equals
\begin{equation}\label{eq:C_j_old}
C_j = n_j \cdot v_\mathrm{rel} \cdot \sigma_\mathrm{col},
\end{equation}
where the collisional cross section equals the geometrical one $\sigma_\mathrm{col}=\pi(s+s_j)^2$, and the relative velocity is a function of both particle sizes and their location in the disk (see Fig. \ref{fig:vels}). The solid red contour in Fig. \ref{fig:vels} marks $v_\mathrm{rel}=v_\mathrm{frag}$, and can be used to identify $s_\mathrm{max}$. Collision velocities are generally higher away from the midplane because the lower gas densities result in higher Stokes numbers for a given size. As a result, the maximum size is a function of height in the disk.

The total collision rate between a single particle of size $s$ and the full dust distribution then becomes
\begin{equation}\label{eq:C_tot}
C_\mathrm{tot} = \sum_j^{N_f} C_j.
\end{equation}
Neglecting the possibility of multiple collisions, the chance of our particle suffering a collision during a period $\Delta t$ equals
\begin{equation}
P_\mathrm{col} = 1- e^{ - C_\mathrm{tot} \Delta t  }.
\end{equation}
Since we do not take into account the possibility of multiple collisions in a single time step, $\Delta t$ should be smaller than the typical collision timescale, or $\Delta t \lesssim 1/C_\mathrm{tot}$. This requirement can become problematic, especially when the followed particle is large and located in the midplane. Large particles will collide frequently with (sub)micron grains, and these collisions will dominate and cause $C_\mathrm{tot}$ to become very large even though the cumulative mass gain/loss resulting from these collisions might well be negligible. 

To avoid forcing the time-step to very small values, background dust particles that are considerably smaller than the followed particle are grouped \citep[e.g.,][]{ormel2008,krijt2015}. The followed particle $s$ encounters these groups less frequently, but when it does, it collides with all the particles in that group. In this work\footnote{See \citet{ormel2008} for a discussion on different grouping strategies.}, we group based on particle mass: in size-bins that correspond to grains with a mass smaller than $f_\epsilon m$ (with $m$ the mass of the aggregate we are following), particles are combined in groups with a total mass $f_\epsilon m$, consisting of $f_\epsilon ( s / s_j)^3$ individual $s_j$-size dust particles. We will use $f_\epsilon=10^{-1}$, so that the mass of the particle with size $s$ cannot change by more than 10\% during one group collision.

\begin{figure*}[!ht]
\centering
\includegraphics[clip=,width=1.\linewidth]{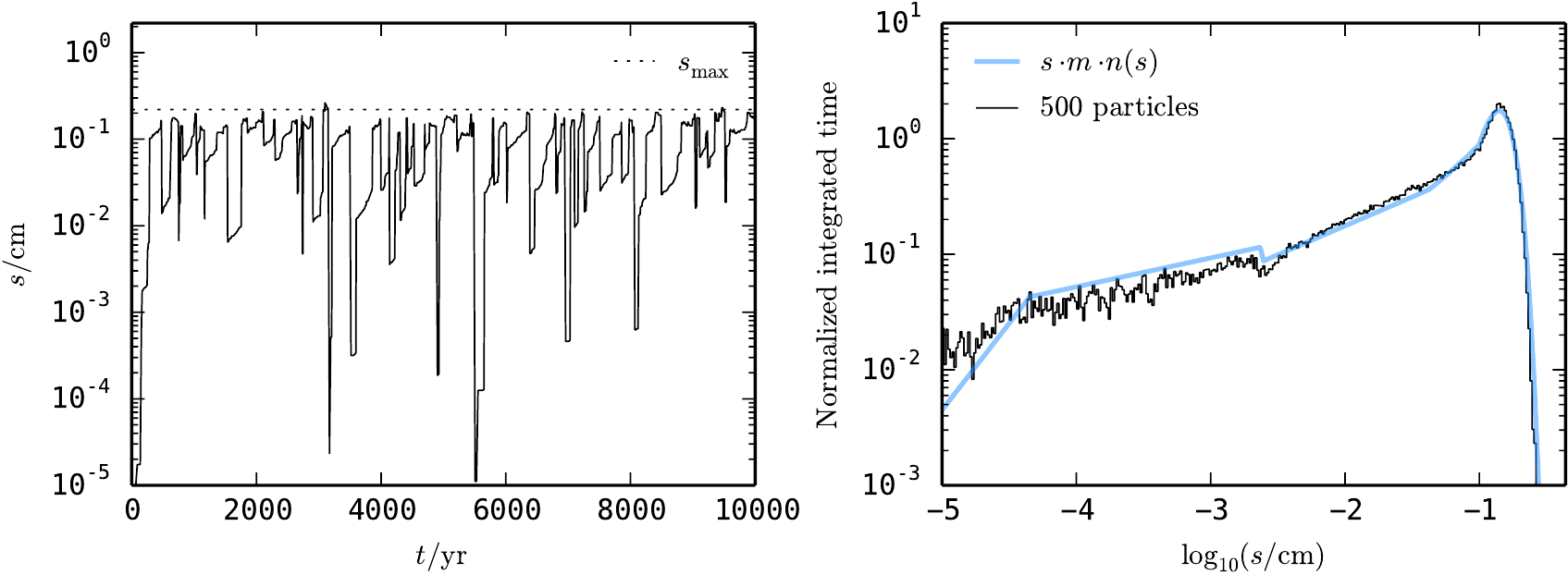}
\caption{History of monomers in a disk with $\alpha=10^{-2}$ in the midplane at $r=5\mathrm{~AU}$ and a local dust-to-gas ratio of 0.01. \emph{Left:} Aggregate sizes in which a single monomer is incorporated as a function of time. During 10,000 years, many sticking, erosion, and fragmentation events occur, and the small grain we are following is part of many different aggregates. \emph{Right:} Normalized histogram of the cumulative time grains spent inside bodies of size $s$, averaged over $500$ monomer histories of 10,000 years each (excluding the first 1000 years). Overall, the histogram is in good agreement with the mass-weighted background size-distribution.}
\label{fig:onlycol}
\end{figure*}

The modified collision rates then become
\begin{equation}\label{eq:C_tilde}
\widetilde{C}_j = \begin{cases}
~C_j &\textrm{~~if~~} (s_j/s)^3 > f_\epsilon,  \vspace{3mm}\\
~f_\epsilon^{-1} (s_j/s)^3 \cdot C_j &\textrm{~~if~~} (s_j/s)^3 \leq f_\epsilon,
\end{cases}
\end{equation}
where the first case describes 1-to-1 collisions between the followed particle $s$ and a single other particle with a similar or much larger size. The second case refers to collisions between the followed particle and \emph{a whole group} of $s_j$-size particles, for which the individual collision rate of Eq. \ref{eq:C_j_old} is modified by a factor corresponding to the number of particles making up the group.

Similar to before, we define
\begin{equation}
\widetilde{C}_\mathrm{tot} = \sum_j^{N_f} \widetilde{C}_j,
\end{equation}
and 
\begin{equation}
\widetilde{P}_\mathrm{col} = 1- e^{ - \widetilde{C}_\mathrm{tot} \Delta t  }.
\end{equation}

To determine if a collision occurred during a time step, we draw a random number $\mathcal{R}_2$ between $[0,1]$, and write
\begin{equation}
\begin{cases}
\textrm{~No collision}  & \textrm{if~~} \mathcal{R}_2 > \widetilde{P}_\mathrm{col}, \vspace{3mm} \\ 
\textrm{~collision} & \textrm{if~~} \mathcal{R}_2 \leq \widetilde{P}_\mathrm{col}.
\end{cases}
\end{equation}
If no collision has taken place, we simply move on to the next time step. If a collision does take place, we need to determine from which size bin the collider was. For this we use the standard approach for drawing random outcomes that follow a distribution (here, $\widetilde{C}_j$): we draw a third random number $\mathcal{R}_3$ between $[0,C_\mathrm{tot}]$ and sum over all bins until we reach bin $k$ for which
\begin{equation}
\sum_{k \leq j}^{N_f} C_k \geq \mathcal{R}_3,
\end{equation}
identifying a particle from bin $k$ as the collision partner. By drawing a fourth random number and comparing it to $P_\mathrm{frag}$, we determine the outcome of the collision in question\footnote{When a collision occurs between a large particle and a group of small particles (the second case in Eq. \ref{eq:C_tilde}), we use the same outcome for all the individual  collisions this event represents. In other words, either \emph{all} small particles stick, or they \emph{all} result in erosion, even if the collisions occur in the transition regime of Eq. \ref{eq:P_frag}.}. Finally, we have to determine in which collision product the followed monomer ends up. In a sticking collision this is straightforward (there is only 1 collision product) while in the case of fragmentation a fifth random number is used to choose an aggregate from the fragment distribution in a mass-weighted manner.

To test the collisional behavior, we run a simulation where the dust grains are confined to the midplane at all time (i.e., $z(t)=0$), for the parameters of Table \ref{tab:benchmark}, but with an increased level of turbulence $\alpha=10^{-2}$ so that $s_\mathrm{max} < s_\mathrm{sett}$ and settling is not important -- even for the largest aggregates. We confine particles to the mid-plane to remove the dependence of the Stokes numbers and collision velocities on height $z$, as these effects were not included in \citet{birnstiel2011}. Figure \ref{fig:onlycol} shows the results for this test using a variable time step $\Delta t = 0.5 /\tilde{C}_\mathrm{tot}$. The left panel shows the history of 1 single monomer. This grain started out with a size $s_\bullet$ at $t=0$, and was left to collide with the background dust population according to the scheme developed in Sect. \ref{sec:colbehav}. During 10,000 years, the grain is part of many different aggregates, as these aggregates suffer collisions that result in sticking, erosion, and fragmentation. Sudden jumps to small sizes indicate fragmentation/erosion events where the grain we are interested in ended up in a small fragment. Conversely, sudden jumps to large sizes indicate our grain is swept up by a much larger particle. There are also periods of more gradual mass increase (for example just before $t=2000\mathrm{~yr}$), indicating that the aggregate we are following is growing by sweeping up smaller particles. Which of these processes (dramatic events or gradual mass gain/loss) dominate can vary per particle size and depend on the velocity field and dust size distribution.

In a steady state, the number density of particles of some size $s$ is set by a balance of their creation and removal: a high abundance means these grains are created frequently and/or survive for a long time, while a low abundance means they are created sporadically and/or are removed quickly after creation. Since we are effectively following single units of mass at $z=0$, and the collision model used for the followed particles is the same as the one used in Sect. \ref{sec:background}, the ergodic principle states that, given enough time (in this case, multiple coagulation/fragmentation cycles), the fractional time spent inside grains of size $s$ should be proportional to the local mass distribution. Moreover, since settling is not important for $\alpha=10^{-2}$, the midplane mass density is directly proportional to the vertically integrated surface density, i.e., $n_i m_i \propto \Sigma_{\mathrm{d},i}$ (Eq. \ref{eq:n_j}). Thus, if we sum over enough monomer histories, we should retrieve the background dust distribution. The right panel of Fig. \ref{fig:onlycol} shows the normalized, cumulative time monomers spent inside a grain of size $s$, averaged over 500 monomer histories, each followed for $10^4\mathrm{~yr}$. The mass-weighted background size distribution (Sect. \ref{sec:background}) is shown by the blue solid line. The curves agree very well, especially towards larger sizes, where most of the mass is located.

\subsection{Combining dynamics and collisions}
When the dynamics and collisions are treated simultaneously, the time step is chosen as
\begin{equation}\label{eq:dt}
\Delta t = \min \left\{ \frac{f_\mathrm{diff}}{(\alpha+\Omega t_s) \Omega} , \frac{f_\mathrm{coll}}{ \widetilde{C}_\mathrm{tot}} \right\},
\end{equation}
and we will typically use $f_\mathrm{diff} \sim 10^{-2}$ and $f_\mathrm{coll} \sim 1$. In this way, vertical diffusion/settling limits the time step when collisions are rare (for example in the upper parts of the disk atmosphere) and the collision rates limit the time step when grains collide more frequently.

The steps for calculating the evolution of a particle that starts out as a monomer inside an aggregate of size $s$ at $t=0$ are then:
\begin{enumerate}
\item{Calculate $\widetilde{C}_j$ and $\widetilde{C}_\mathrm{tot}$ for the particle of size $s$ at position $z$.}
\item{Determine the time step $\Delta t$ using Eq. \ref{eq:dt}.}
\item{Displace the grain vertically according to Eq. \ref{eq:z_new}.}
\item{Determine if a collision has occurred during the time step $\Delta t$.}
\item{If yes, determine the collision partner, calculate the collisional outcome (sticking, fragmentation, or erosion).}
\item{Given the collisional outcome, determine the size of the body the monomer we are following ends up in.}
\item{Repeat all steps while $t < t_\mathrm{max}$.}
\end{enumerate}
Typically we will take $t_\mathrm{max} \sim 10t_D$, where $t_D = 1/(\alpha \Omega) \simeq 1.7 \cdot10^{3}\mathrm{~yr}$ for the parameters of Table \ref{tab:benchmark}.

\section{Individual histories}\label{sec:individual}
\begin{figure}
\centering
\includegraphics[clip=,width=1.\linewidth]{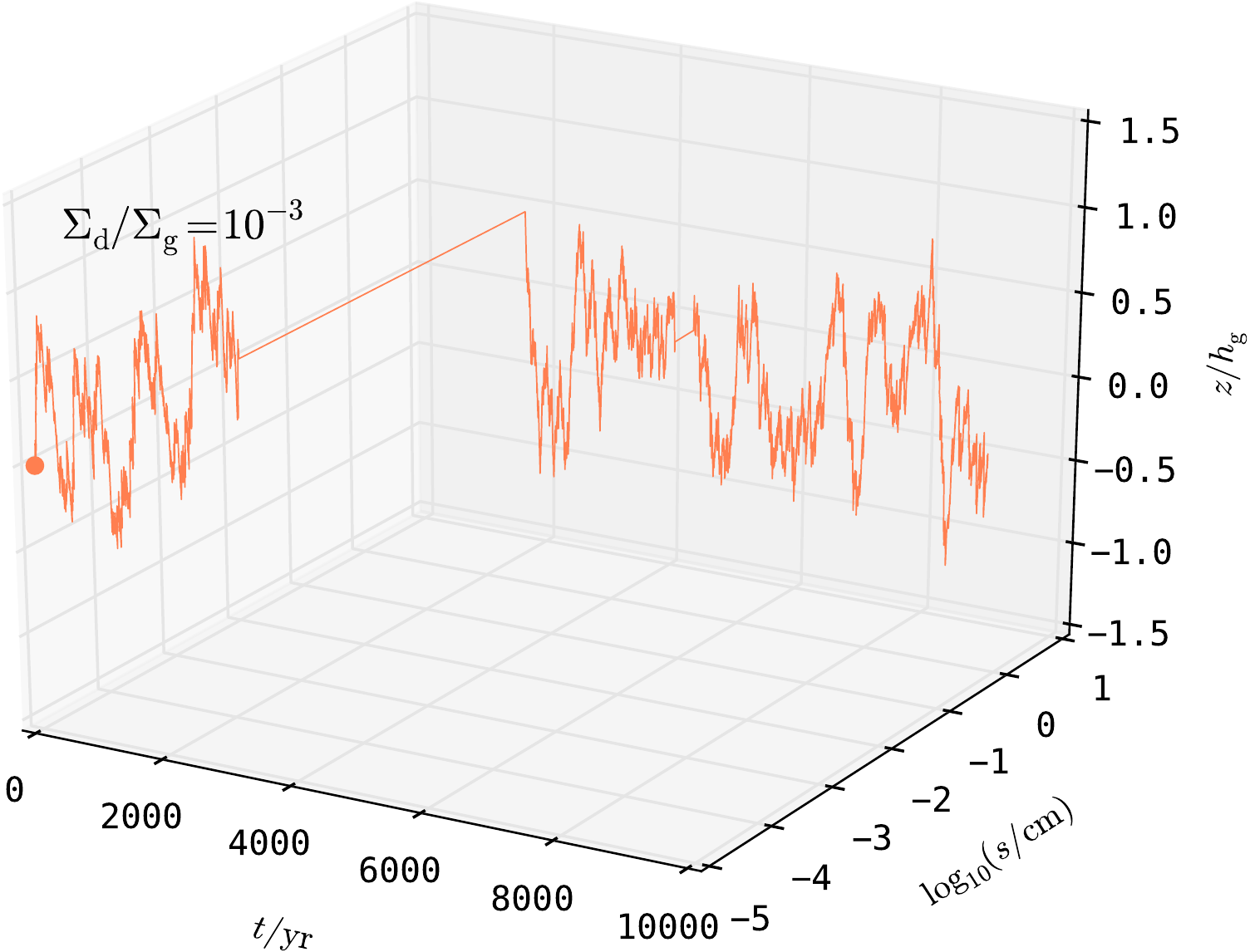}\\ \vspace{1mm}
\includegraphics[clip=,width=1.\linewidth]{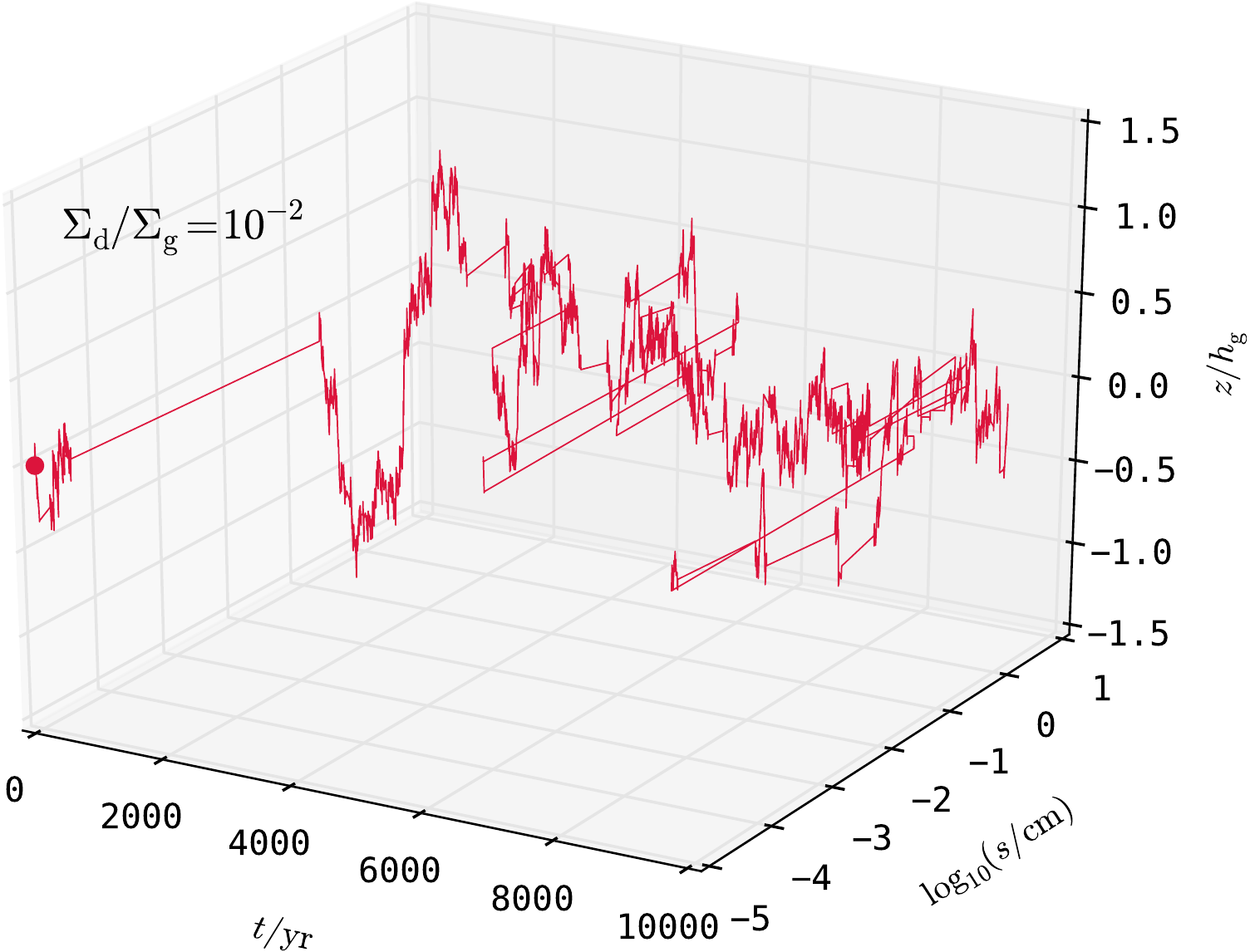}\\ \vspace{1mm}
\includegraphics[clip=,width=1.\linewidth]{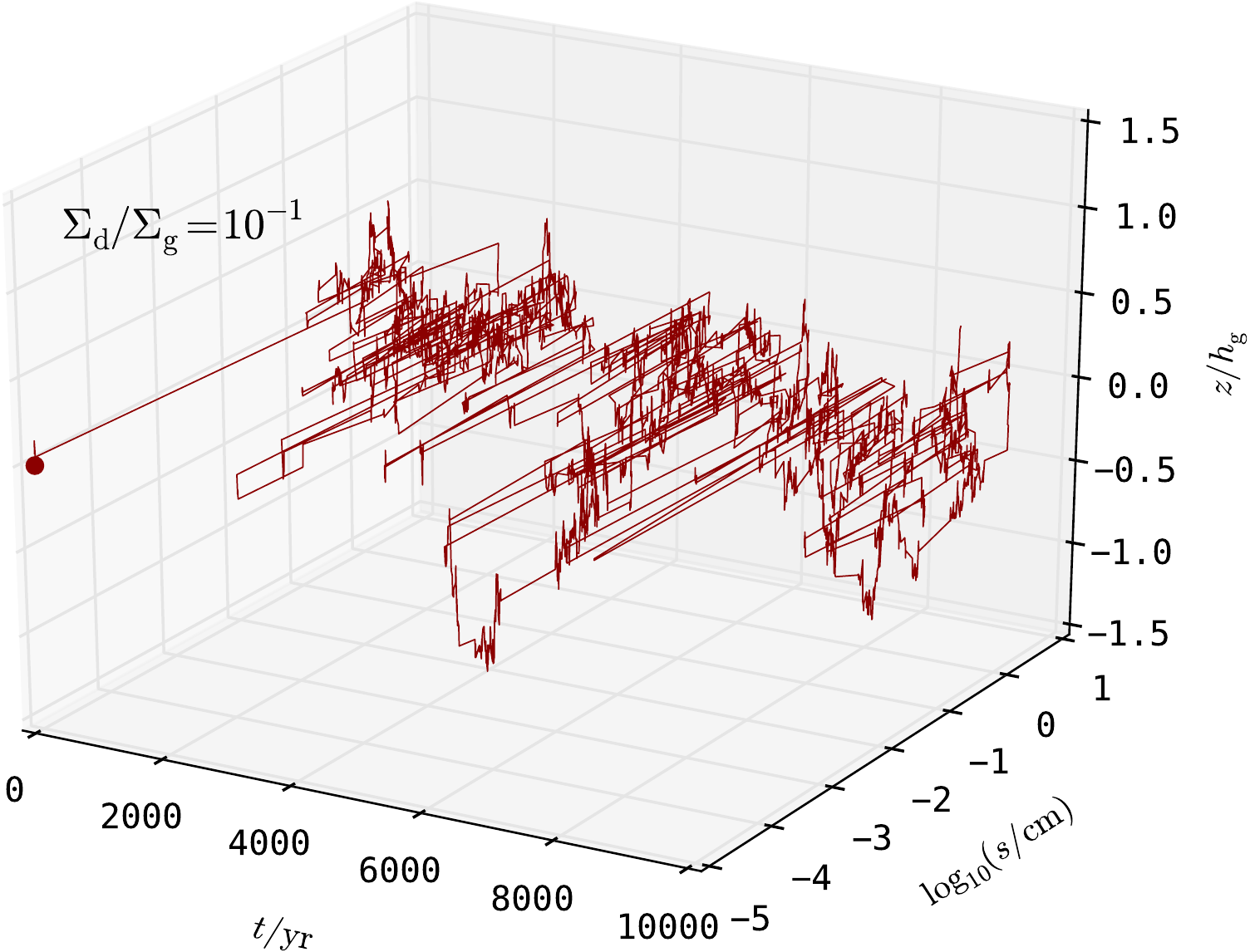}
\caption{Histories of monomers starting out in the midplane at $r=5\mathrm{~AU}$ in a disk with $\alpha=10^{-3}$ for three different dust-to-gas ratios. The dust population has the shape of the one in Fig. \ref{fig:nama}, but the total dust mass is varied. After they are released at the location of the $\bullet$-symbol, particles can diffuse and settle vertically, and are incorporated in different aggregates as the result of collisions.}
\label{fig:3d}
\end{figure}

To illustrate the interplay between diffusion and particle-particle collisions, we calculate individual histories for monomer grains released in the midplane at $5\mathrm{~AU}$ for three different dust-to-gas ratios $\Sigma_\mathrm{d}/\Sigma_\mathrm{g}=10^{-3},10^{-2},10^{-1}$ corresponding to dust depleted, Solar, and dust enhanced ratios, respectively. The turbulence strength $\alpha=10^{-3}$ and the shape of the background dust population is identical to the one shown in Fig. \ref{fig:nama}. Independent of $\Sigma_\mathrm{d}/\Sigma_\mathrm{g}$, the maximum particle size equals $s_\mathrm{max}\simeq2\mathrm{~cm}$ and has a (midplane) Stokes number of $\Omega t_s \simeq 0.02$. Each panel of Fig. \ref{fig:3d} shows the location and size of the aggregate the monomer is part of as a function of time.

For the lowest dust density (upper panel of Fig. \ref{fig:3d}), collisions are not very frequent. After the monomer is released at $z=0$, it moves vertically for a few thousand years before being swept up by a cm-size aggregate. Within the simulated 10,000 years, only a handful of collisions have occurred and the monomer stays part of the same grain for extended periods of time. In the middle panel, where the dust density is 10 times higher and at the Solar value, this picture changes. In the same 10,000 years, this monomer witnesses multiple sticking and destructive collisions and as a result spends more time inside aggregates of a variety of sizes. Finally, in the dust-rich case where $\Sigma_\mathrm{d}/\Sigma_\mathrm{g}=10^{-1}$ (lower panel), collisions happen very frequently and the monomer we are following rarely stays part of the same aggregate for an extended period of time. 

In summary, a higher $\Sigma_\mathrm{d}/\Sigma_\mathrm{g}$ means collisions play a larger role in shaping the dynamical evolution of the individual dust grains. In the next section, we study how frequent collisions can influence the vertical (re)distribution of particles.

\section{Trapping of small grains}\label{sec:collective}
In the growth/fragmentation steady state considered here, small (sub)micron particles are created predominantly in catastrophic collisions between particles with sizes close to the maximum size $s_\mathrm{max}$ because \emph{i)} this is where most of the solid mass is concentrated (Fig. \ref{fig:onlycol}), and \emph{ii)} relative velocities between smaller particles result in sticking instead of fragmentation (Fig. \ref{fig:vels}), allowing growth to replenish the population of $s_\mathrm{max}$ particles.

If these larger bodies are concentrated in a thin layer because of settling, the production of small grains will be highest inside that layer. With particle collisions happening very frequently in some cases (see Fig. \ref{fig:3d}), it is possible that in some scenarios fragments will not survive long enough to be redistributed vertically before being swept up by the next aggregate. In this section, the impact of coagulation/fragmentation on the vertical distribution of small grains is explored for a dust column at $5\mathrm{~AU}$ for different dust contents and turbulence strengths.

\begin{figure*}[t]
\centering
\includegraphics[clip=,width=1.\linewidth]{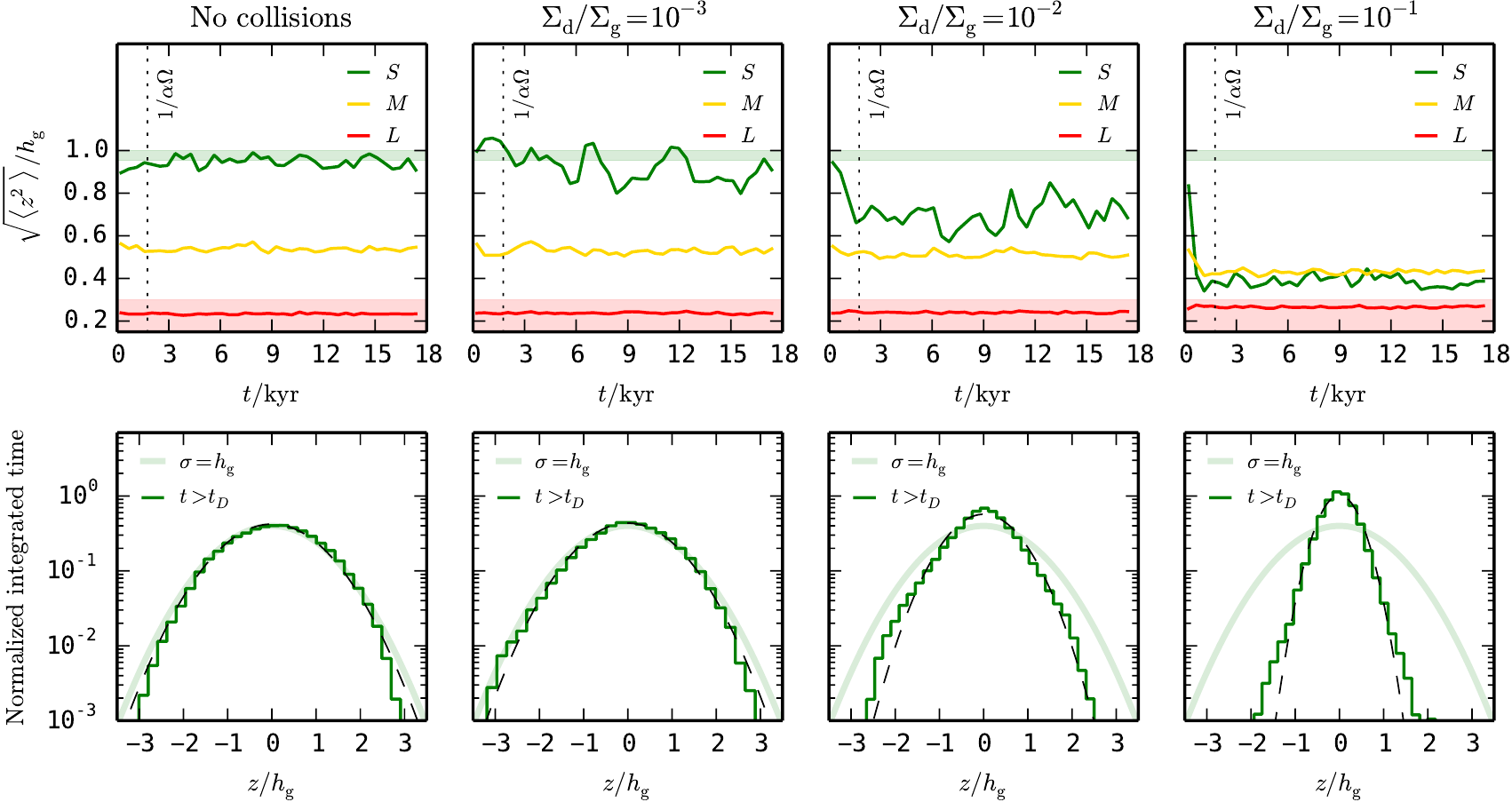}
\caption{Dust particle distributions for $\alpha=10^{-3}$ for different dust-to-gas ratios. \emph{Top:} Variance in the distribution of small ($\Omega t_s/\alpha < 10^{-1}$), medium ($10^{-1}<\Omega t_s/\alpha < 10$) and large ($\Omega t_s / \alpha > 10$) particles as a function of time. The shaded areas indicate the expected scale heights for small and large grains (i.e., Eq. \ref{eq:h_d}), and the vertical dotted line shows the mixing timescale $t_D=1/(\alpha \Omega)$. \emph{Bottom:} Normalized distribution of small grains integrated over all times $t>t_D$. The shaded green curve shows the expected distribution for well-mixed small grains and the dashed line shows a Gaussian fit to the data.}
\label{fig:vars}
\end{figure*}

\subsection{Varying the dust content}\label{sec:varying_sigma}
Here, we start by recreating the steady-state distribution (Sect. \ref{sec:background}) as closely as possible. Specifically, the monomers are distributed inside aggregates with initial sizes $s_i(t=0)$ drawn from the steady-state size distribution and given initial locations $z_i(t=0)$ drawn from Gaussians centered at $z=0$ with widths $h_\mathrm{d}$ corresponding to $s_i$. Then, all grains are followed over a time period $t=10t_D$ while they diffuse, settle, and become part of different aggregates through collisions resulting in sticking, catastrophic fragmentation and erosion. In total, $10^3$ individual monomer histories (similar to those shown in Fig. \ref{fig:3d}) are calculated. Depending on their midplane Stokes number, aggregates are classified as being \emph{small} ($\Omega t_s/\alpha < 10^{-1}$), \emph{intermediate} ($10^{-1}<\Omega t_s/\alpha < 10$), or \emph{large} ($\Omega t_s / \alpha > 10$). For each group, their scale-height is then obtained by calculating the root-mean-square of their distribution, i.e., $h_\mathrm{d}=  \langle z_i^2 \rangle ^{1/2}$, assuming that $\langle z_i \rangle=0$.

Figure \ref{fig:vars} shows the results for $\alpha=10^{-3}$ and \emph{four} different dust-to-gas ratios. The top panels show the evolution of the scale-height of the small, intermediate, and large grains as a function of time. The green and red shaded areas indicate $h_\mathrm{d}$ for small and large particles as predicted by Eq. \ref{eq:h_d}, and the vertical dotted line corresponds to the mixing time $t_D$ (Eq. \ref{eq:t_mix}). Bottom panels show the distribution of small particles summed over all times $t>t_D$. The dashed line is a Gaussian fit to the observed distribution, and the green shaded curve shows the distribution of the gas. Dust-to-gas ratios increase from left-to-right with the left-most column showing the result of simulations without taking into account collisions (effectively $\Sigma_\mathrm{d}/\Sigma_\mathrm{g} \rightarrow0$). 

Starting on the left, it can be seen that grains behave as expected in the absence of collisions (see also Fig. \ref{fig:noncol}). When the total dust mass is increased however, the impact of successive growth/fragmentation cycles on the vertical distribution of grains is clearly visible. The effect is largest for the small grains, whose average scale-height decreases from almost 1 to ${\sim}0.4$, a value closer to the scale-height of the large particles in the steady-state population (see Table \ref{tab:dustpops}). The small particles that are created in the midplane cannot move freely for long enough to reach the upper parts of the disk.

\subsection{Varying the turbulence}
It is interesting to study how the results vary with the strength of the turbulence. Changing $\alpha$ will impact the behavior in many ways, influencing (among other things) the shape and upper limit of the size-distribution, the settling behavior, and the diffusion timescale. Table \ref{tab:dustpops} shows how some important aspects of the model vary with $\alpha$.

\begin{table*}[!ht]
\begin{center}
\caption{Characteristics of normalized dust populations at $r=5\mathrm{~AU}$ for $v_\mathrm{frag}=5\mathrm{~m~s^{-1}}$ for different values of $\alpha$. \label{tab:dustpops}      }
\begin{tabular}{c  c c c c | c c c | c c}
\tableline\tableline
$\alpha$  & $s_\mathrm{sett}/\mathrm{cm}$ & $s_\mathrm{p}/\mathrm{cm}$ & $(\Omega t_s)_\mathrm{p}$ & $(h_\mathrm{d}/h_\mathrm{g})_\mathrm{p}$ &   $\Delta^{(S)}$ & $\Delta^{(M)}$ & $\Delta^{(L)}$ & $t_D/\mathrm{yr}$ & $t_\mathrm{drift} /\mathrm{yr}$\\
\tableline
$10^{-2}$   &  $0.81$ & $0.140$ & $0.0017$ & $0.92$ & $0.521$ & $0.479$ & $0$ & $1.8 \cdot10^{2}$ & $2.7\cdot10^{5}$ \\

$3\cdot10^{-3}$ &  $0.24$ & $0.468$ & $0.0058$ & $0.58$ & $0.184$ & $0.816$ & $0$ & $5.9\cdot10^{2}$ & $6.2\cdot10^{4}$ \\

$10^{-3}$ &   $0.081$ & $1.40$ & $0.017$ & $0.23$ & $0.116$ & $0.491$ & $0.393$ & $1.8\cdot10^{3}$ & $2.7\cdot10^{4}$ \\

$3\cdot10^{-4}$ & $0.024$ & $4.68$ & $0.058$ & $0.07$ & $0.073$ & $0.251$ & $0.675$ & $5.9\cdot10^{3}$ & $6.2\cdot10^{3}$ \\

 \tableline
\end{tabular}
\tablecomments{From left to right, the table lists: the turbulent $\alpha$; the size of the particles that begin to settle; the size, Stokes number, and scale-height (through Eq. \ref{eq:h_d}) of the mass dominating particles; the normalized mass $\Delta$ in small, intermediate, and large particles (see Sect. \ref{sec:varying_sigma} for the definitions of these sizes); the mixing timescale; and the drift timescale of particles with size $s_\mathrm{p}$ (see Appendix \ref{sec:lifetime}). All quantities listed in this table are insensitive to the choice of $\Sigma_\mathrm{d}/\Sigma_\mathrm{g}$.}
\end{center}
\end{table*}

\begin{figure}[!ht]
\centering
\includegraphics[clip=,width=1.\linewidth]{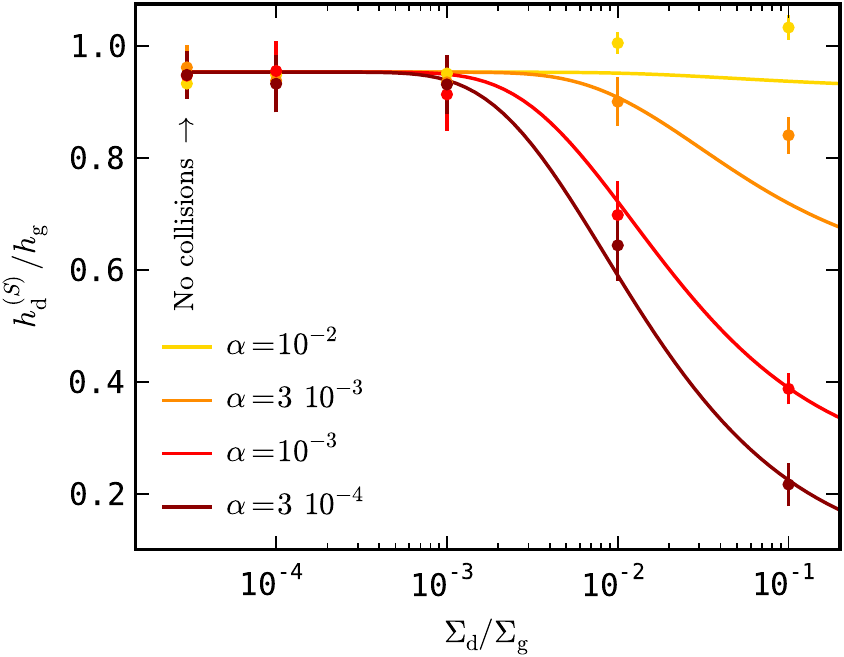}
\caption{Scale height of the small dust grains (those with $\Omega t_s < \alpha/10$). Points shows the results of simulations like the ones shown in Fig. \ref{fig:vars} and lines represent Eq. \ref{eq:fit}.}
\label{fig:hd_small}
\end{figure}

In Fig. \ref{fig:hd_small} the results for a series of simulations with different turbulence strengths have been summarized. For every value of $\alpha$, the dust-to-gas ratio has been varied as in Sect. \ref{sec:varying_sigma}. The value of $h_\mathrm{d}^{(S)}$ has been obtained by averaging $10^3$ particle histories in the time frame $t_D \leq t \leq 10 t_D$. The vertical lines indicate the variance $h_\mathrm{d}^{(S)}$ exhibits when binned in 36 periods of $0.25t_D$ each. The left-most points correspond to simulations where collisions were ignored completely (like in Sect. \ref{sec:diffbehav}). In all simulations with dust-to-gas ratios $\Sigma_\mathrm{d}/\Sigma_\mathrm{g} \gtrsim 10^{-3}$, small grains are trapped in the midplane. The decrease of $h_\mathrm{d}^{(S)}$ is most pronounced for low values of $\alpha$, where the largest grains have settled most (Table \ref{tab:dustpops}) and the mixing timescale is longer (Eq. \ref{eq:t_mix}). 

For $\alpha=10^{-2}$, our simulations indicate $h_\mathrm{d}^{(S)}>1$ which appears unphysical. This is the result of increased small dust production at high $z$ because the collision velocity is increased with respect to the midplane (see both panels of Eq. \ref{fig:vels}). This broadens the vertical distribution of small grains slightly, resulting in $  \langle z_i^2 \rangle ^{1/2}  > h_\mathrm{g}$. This effect is much smaller in the simulations with a lower $\alpha$, where almost no destructive collisions take place at high $z$ because most of the mass is in the midplane. We return to a discussion of how the assumptions of our model may impact our results further below.

\subsection{Comparing timescales}
The trapping of small particles seen here can be understood by comparing the mixing timescale to the survival time of small grains near the midplane, where they are formed. When the removal of small grains from the midplane is dominated by collisions with particles of size $s_\mathrm{p}$ and Stokes number $(\Omega t_s)_\mathrm{p}$, the sweep-up timescale can be written as
\begin{align}
t_\mathrm{sw} &= \left( n_\mathrm{p} \sigma_\mathrm{col} v_\mathrm{rel} \right)^{-1},\\
&\simeq \dfrac{h_\mathrm{g}}{c_s \Sigma_\mathrm{d}}  \dfrac{ s_\mathrm{p} \rho_\bullet}{ (\Omega t_s)_\mathrm{p}} , \label{eq:t_sw_1}
\end{align}
where we have assumed the collision velocity $v_\mathrm{rel} \sim c_s \sqrt{\alpha (\Omega t_s)_\mathrm{p}}$ is set by the larger particles (see Fig. \ref{fig:vels}), and approximated the number density of peak-mass particles near the mid-plane $n_\mathrm{p} \sim \Sigma_\mathrm{d}/(h_\mathrm{d} m_\mathrm{p} )$ with $h_\mathrm{d} / h_\mathrm{g} \sim \sqrt{\alpha / (\Omega t_s)_\mathrm{p}}$. Assuming the peak-mass particles are in the Epstein regime (Eq. \ref{eq:t_s}) and making use of $\rho_\mathrm{g} \sim \Sigma_\mathrm{g} / h_\mathrm{g}$, the particle size and material density drop out and we are left with the surprisingly simple relation
\begin{equation}
t_\mathrm{sw} \sim \frac{1}{\Omega}  \left( \frac{\Sigma_\mathrm{d}}{\Sigma_\mathrm{g}}\right)^{-1}.
\end{equation}

It is interesting to compare the sweep-up timescale to the mixing time of the smallest particles $t_D$ (Eq. \ref{eq:t_mix}). We define the ratio of these two timescales
\begin{equation}\label{eq:Phi}
\Phi \equiv \frac{t_\mathrm{sw}}{t_D} \sim \alpha \left( \frac{\Sigma_\mathrm{d}}{\Sigma_\mathrm{g}}\right)^{-1}.
\end{equation}
For $\Phi \gg 1$, small grains can easily diffuse before being swept up, and we can expect to retrieve $h_\mathrm{d}^{(S)} = h_\mathrm{g}$. Alternatively, when $\Phi \ll 1$, the sweep-up time is short compared to the mixing timescale. In this limit, we expect $h_\mathrm{d}^{(S)}$ to decrease and approach the scale height of the mass-dominating particles. From our simulations, we find that the transition between these regimes occurs around $\alpha = \Sigma_\mathrm{d}/\Sigma_\mathrm{g}$. This picture and the location of the regime boundary are consistent with the behavior seen in Figs. \ref{fig:3d} and \ref{fig:vars}.

Finally, it is useful to construct a function that recreates this behavior and provides the scale height of small ($\Omega t_s<\alpha/10)$ particles. We find that
\begin{equation}\label{eq:fit}
\frac{ h_\mathrm{d}^{(S)} }{h_\mathrm{g}} = ( y_2-y_1 ) e^{ - ( c_1 \Phi  )^{2/3}} + y_1,
\end{equation}
with $y_1 \simeq 0.95 $ the relative scale-height of particles with $\Omega t_s = \alpha/10$ (the largest of the small particles) and $y_2$ the relative scale height of the mass-dominating particles (listed in Table \ref{tab:dustpops}) works well in most cases. In Figure \ref{fig:hd_small}, the predictions of Eq. \ref{eq:fit} are plotted alongside with the results from our simulation. Fitting Eq. \ref{eq:fit} to these results suggests $c_1=16$ for $\alpha=3\cdot10^{-4}$ and $c_1=7$ for more turbulent cases.

\section{Discussion}\label{sec:disc}

\subsection{Evaluating model assumptions}\label{sec:validity}
A key assumption that is made in this study is that the background dust population is well-described by a growth/fragmentation steady state distribution as described by \citet{birnstiel2011} (see Sect. \ref{sec:background}). This implies that the maximum particle size is set by turbulence-induced fragmentation. For this to be the case, the relative collision velocity should exceed the fragmentation threshold, i.e., $\Delta v_\mathrm{tur} \sim \sqrt{ \alpha } c_s < v_\mathrm{frag}$. This is usually achieved in the inner disk, where $c_s$ is higher. In addition, bare silicate grains inside the snow-line have a significantly lower fragmentation threshold than their icy cousins in the outer disk \citep[e.g.,][]{wada2013}. Second, growth to sizes ${\sim}s_\mathrm{max}$ has to be able to proceed unhindered by effects such as bouncing or radial drift. The latter is often problematic in the outer disk, where particles drift radially before reaching $s_\mathrm{max}$, resulting in a maximum size that can be significantly smaller \citep{brauer2007,birnstiel2010,birnstiel2012,estrada2016}.

Even when growth up to $s_\mathrm{max}$ is possible, the most massive particles will drift radially and the local dust surface density will typically decrease. The time-scale on which this occurs depends on the Stokes numbers of the largest particles, and can be estimated for the calculations presented here. The last two columns of Table \ref{tab:dustpops} show $t_D$ and $t_\mathrm{drift}$ (see Appendix \ref{sec:lifetime}). While a stronger turbulence decreases the mixing time, it increases the drift timescale because the maximum particle size and corresponding Stokes number are lower (e.g., Eq. \ref{eq:s_max}). In all cases considered here, $t_\mathrm{drift} \gg t_D$ except for $\alpha=3\cdot10^{-4}$ where the timescales are comparable.

Thus, the simulations presented in this work are probably most relevant for the inner parts of protoplanetary disks, where growth and mixing are fast, collision velocities are high, and the fragmentation threshold is lower. Lastly, throughout this work we have assumed that the largest particles are always in the Epstein drag regime. In the very inner disk, or for very porous aggregates, this might not always be true. In those cases, the Stokes drag regime applies, in which the stopping time is increased by a factor $s/\lambda_\mathrm{mfp} > 1$ \citep{cuzzi2006}. Inserting this factor in Eq. \ref{eq:t_sw_1} results in a \emph{shorter} sweep-up time (by a factor $\lambda_\mathrm{mfp}/s_\mathrm{p}$), making collisional trapping more effective.

\subsection{Small dust at high altitudes}\label{sec:high_z}
A smaller scale height for the smallest dust particles has a major impact on their number density in the upper disk atmosphere. The expected local small-dust-to-gas ratio at a height $z$ can be written as
\begin{equation}\label{eq:local}
\frac{\rho_\mathrm{d}^{(S)}}{\rho_\mathrm{g}} = \Delta^{(S)}\frac{\Sigma_\mathrm{d}}{\Sigma_\mathrm{g}} \left( \frac{h_\mathrm{d}}{h_\mathrm{g}}\right)^{-1} \exp \left\{ -\frac{1}{2}\left( \frac{z^2}{h_\mathrm{d}^2} - \frac{z^2}{h_\mathrm{g}^2} \right)  \right\},
\end{equation}
with $h_\mathrm{d}=h_\mathrm{d}^{(S)}$ the scale-height of the small dust and $\Delta^{(S)}$ the fraction of the size-distribution residing in small grains (see Table \ref{tab:dustpops}).

\begin{figure}[t]
\centering
\includegraphics[clip=,width=1.\linewidth]{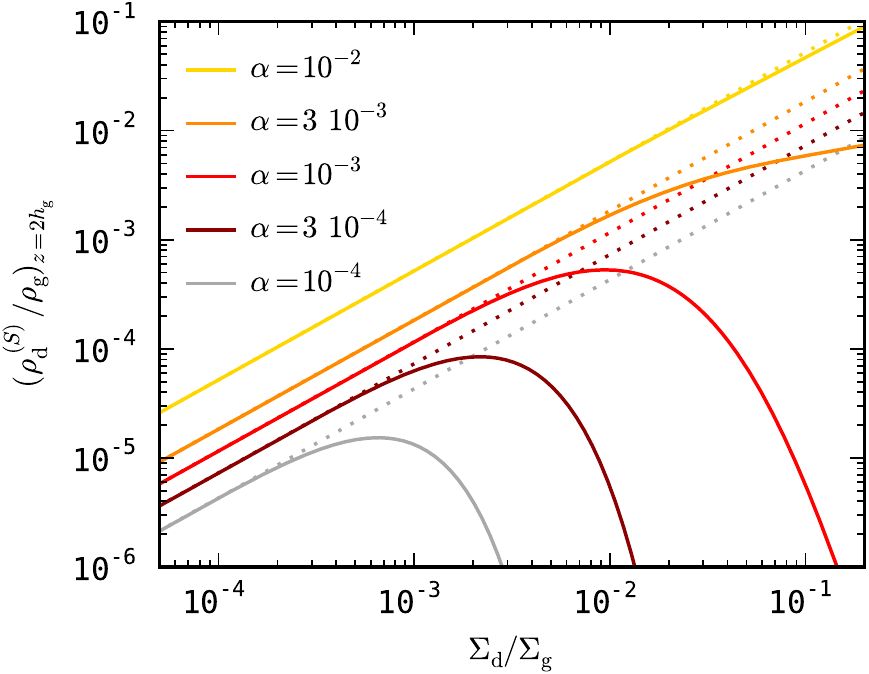}
\caption{Expected local small-dust-to-gas ratio at $z=2h_\mathrm{g}$ (Eq. \ref{eq:local}) for the parameters of Table \ref{tab:benchmark} and different turbulence levels and $\Sigma_\mathrm{d}/\Sigma_\mathrm{g}$. Dotted lines assume the small dust is well-mixed ($h_\mathrm{d}^{(S)}=h_\mathrm{g}$) and solid lines assume the scale-height of small grains is set by collisional trapping and accurately described by Eq. \ref{eq:fit} using $c_1=10$ and $y_1=1$.}
\label{fig:smalldust}
\end{figure}

Figure \ref{fig:smalldust} shows the dust-to-gas ratio at $z=2 h_\mathrm{g}$ as a function of the total vertically-integrated dust-to-gas ratio. The dotted lines show the predictions for the cases where small grains are always well-mixed and $\rho_\mathrm{d}^{(S)}/\rho_\mathrm{g} \propto \Sigma_\mathrm{d}/\Sigma_\mathrm{g}$. Differences here are because of different values of $\Delta^{(S)}$. The solid lines show predictions for when collisional trapping taken into account by using Eq. \ref{eq:fit}, using $c_1=10$ and $y_1=1$. The turn-over of these curves is interesting and perhaps counter-intuitive: while adding more dust to the column is still increasing the \emph{total} amount of small dust (i.e., we keep $\Delta^{(S)}$ constant), it is also increasing the degree of collisional trapping and thus lowering $h_\mathrm{d}^{(S)}$. Since the local number density scales exponentially with the square the relative scale-height, the latter effect is strongest and the net amount of small dust in the upper atmosphere decreases.

\subsection{Impact on vertically-integrated size distribution}
In this work we have assumed that the vertically integrated dust distribution is given by the fit function of \citet{birnstiel2011} (see Sect. \ref{sec:background}). However, that approach in turn assumes that dust particles are at all times distributed vertically according to Eq. \ref{eq:h_d}. When $\Phi \ll 1$, our models show significant deviations from this pure mixing/diffusion equilibrium and this is expected to impact the size-distribution itself. For example, if small grains are confined to the midplane, their collision rate with large particles will increase, probably reducing their overall abundance. The greater density of solids at the disk midplane will likely lead to more frequent growth than assumed here, and thus our conclusions would be strengthened.

Ideally one would calculate the background distribution self-consistently. \citet{zsom2011} accomplished this by stacking a series of boxes on top of each other, using the representative Monte Carlo model of \cite{zsom2008} to simulate particle coagulation in each box individually, and allowing dust grains to move between adjacent boxes. Focusing on small grains, \citet{zsom2011} conclude that ``high values of turbulence are needed to explain why disk atmospheres are dusty for ${\sim}10^{6}\mathrm{~yr}$''. While it is difficult to compare our results directly, mainly because \citeauthor{zsom2011} used a more complex model for the collisional outcomes \citep{guttler2010} and porosity evolution \citep{ormel2007,okuzumi2009}, our findings are in agreement with their conclusion and, based on the results presented here, we can offer a second possibility, namely low values of $\Sigma_\mathrm{d}/\Sigma_\mathrm{g}$ that allow small grains to diffuse more freely. Such conditions may be realized as solids are depleted by radial drift or as much of the mass gets locked up in planetesimals, unable to take part in the collisional evolution considered here.

While solving the size-distribution in a self-consistent manner is beyond the scope of this work, we can use the framework of \citet{birnstiel2011} to estimate what the size-distribution would look like if all grain sizes have the same scale-height (see Appendix \ref{sec:nama_diff}). From Fig. \ref{fig:nama_diff} we see that collisional trapping will indeed reduce the overall abundance of small grains. The differences between the trapped and non-trapped distributions can be a factor of a few and largest for weak turbulence, where $s_\mathrm{sett}$ and $s_\mathrm{max}$ differ by several orders of magnitude.

\subsection{Signs of radial drift or planetesimal formation?}
Finally, we should discuss inferences that might be made from \emph{not} seeing any dust trapping, i.e., witnessing significant amounts of small dust high in the disk atmosphere. In the context of Eq. \ref{eq:fit}, two things could be happening. First, if the largest bodies in the size distribution do not experience strong settling (i.e., $y_1\sim y_2 \sim 1$ in Eq. \ref{eq:fit}), the trapping is not effective. This happens when $(\Omega t_s)_\mathrm{p} < \alpha$, which can occur either because fragmentation limits growth too early (specifically, when $(v_\mathrm{frag}/c_s) < \alpha$), or when growth is limited by radial drift (Sect. \ref{sec:validity}). Second, a high value of $\Phi$ indicates small grains can move relatively unhindered by collisions, irrespective of whether the large grains have settled or not. For the turbulence strengths considered here, $\Sigma_\mathrm{d}/\Sigma_\mathrm{g}<10^{-3}$ did not see any trapping (see Fig. \ref{fig:hd_small}). For typical initial dust-to-gas ratios of ${\sim}10^{-2}$, this means a reduction of at least one order of magnitude, which could point towards a decrease in the overall dust content caused by radial drift \citep[e.g.,][]{ krijt2015b}, or perhaps indicate that a substantial portion of the dust mass has been converted into planetesimals and no longer takes part in the collisional cascade \citep[e.g.,][]{najita2014}. Finally, the observed small particles might not come from the midplane, but be replenished from outside the disk \citep{dominik2008}. This would suggest no genetic relationship between the dust observed at the disk surface and that which is accreted into planetesimal at disk midplane.

\subsection{TW Hya}
The protoplanetary disk around TW Hya could be an interesting environment to look for observational signatures of collisional dust trapping. First, there exist independent measurements for the dust mass through (sub)-mm observations \citep[e.g.,][]{andrews2012} and the gas mass through rotational HD lines \citep{bergin2013}, which allow us to constrain $\Sigma_\mathrm{d}/\Sigma_\mathrm{g}$. Moreover, there are direct and indirect constraints on the strength of the turbulence \citep{hughes2011,menu2014}. For the inner disk (the region around $2.5\mathrm{~AU}$), \citet{menu2014} argue that the dust population is fragmentation-limited by comparing the inferred radial profile of the dust surface density to the theoretical models of \citet{birnstiel2012}. For the outer disk, several studies have found that the upper layers of the disk's atmosphere are significantly depleted in C and O \citep{du2015,kama2016}. A potential explanation put forward by both groups is that these species have been locked up inside large, settled grains, and are not efficiently being mixed back up. The vapor in the atmosphere can be replenished by small grains diffusing upward (small grains are the prime carriers because of their high surface-to-mass ratio). Thus, when even the smallest grains cannot easily return to the atmosphere, as shown in our simulations, the depletion of volatiles from the atmosphere could be very efficient. While the simulations in this work are not directly applicable to the outer disk (radial drift is expected to play a major role in these parts of the disk), similar modeling can be used to study the  impact of dust dynamics on the volatile budget.

The next step in modeling such a system would be to expand the approach developed here with a description of freeze-out and sublimation of important volatile species and study the distribution of gas and dust in time simultaneously. In low temperature high density regions like the midplane gas molecules will freeze-out onto dust grains, while the molecules are expected to desorb and return to the gas phase in the warmer disk atmosphere \citep[e.g.,][]{ros2013}. Simulating the interplay between these processes, while solving the dust size-distribution in the presence of radial drift self-consistently, will be the subject of future work.


\section{Conclusions}\label{sec:concl}
We have modeled the vertical diffusion and settling behavior of dust grains in a single column of gas in a protoplanetary disk while taking into account sticking and destructive collisions. We assume that the local size distribution is in a growth-fragmentation equilibrium and that radial drift is not an important factor over the $10^3-10^4\mathrm{~yr}$ that we consider. Our models indicate that:

\begin{enumerate}

\item{Contrary to what is usually assumed, small (sub)micron grains are not necessarily well-mixed with the gas, but can become trapped in a layer that is significantly thinner.}

\item{Collisional trapping occurs when the time-scale on which small particles are swept up by larger grains is much shorter than the mixing timescale (see Fig. \ref{fig:vars}). For the steady-state dust populations of Sect. \ref{sec:background}, this occurs when $\Phi \equiv \alpha / (\Sigma_\mathrm{d}/\Sigma_\mathrm{g}) < 1$.}

\item{The scale-height of the small grains (those with Stokes numbers $\Omega t_s < \alpha/10$) ranges from $h_\mathrm{d}\sim h_\mathrm{g}$ (no trapping) to close to the scale-height of the mass-dominating particles (efficient trapping) and is accurately described by Eq. \ref{eq:fit} with $c_1 \sim 10$ (see Fig. \ref{fig:hd_small})}

\item{A reduced scale-height for small dust grains has a large impact on the local dust abundance in the disk atmosphere. Since collisional trapping is more efficient for high $\Sigma_\mathrm{d}/\Sigma_\mathrm{g}$, this leads to the counter-intuitive behavior where \emph{increasing} the column's dust content \emph{decreases} the amount of small dust at high $z$ (Fig. \ref{fig:smalldust}).}

\end{enumerate}

\noindent Future studies are encouraged to investigate the importance of collisional trapping in evolving dust populations. 

\acknowledgments
The authors thank C.\,W.~Ormel, T.~Birnstiel, and M.~Kama for encouraging discussions, and the anonymous referee for his/her insightful comments. This material is based upon work supported by the National Aeronautics and Space Administration under Agreement No. NNX15AD94G for the program ``Earths in Other Solar Systems''. The results reported herein benefitted from collaborations and/or information exchange within NASA's Nexus for Exoplanet System Science (NExSS) research coordination network sponsored by NASA's Science Mission Directorate. The authors acknowledge funding from NASA grants NNX12AD59G and NNX14AG97G.

\appendix

\section{Lifetime of the steady-state dust population}\label{sec:lifetime}
The lifetime of the dust population is set by the drift timescale of the largest particles
\begin{equation}
t_\mathrm{drift} \equiv \frac{r}{v_\mathrm{drift}}.
\end{equation}
If the dominant particles have $\Omega t_s < 1$ the drift velocity can be written as
\begin{equation}
v_\mathrm{drift} =  \eta v_\mathrm{K} \Omega t_s,
\end{equation}
where $v_\mathrm{K}=r\Omega$ and  $\eta \approx (c_s/v_\mathrm{K})^2\simeq0.002$ for the disk model we use. Table \ref{tab:dustpops} shows the drift timescale for the particles making up the peak of the mass distribution. On this timescale, the dust surface density is expected to decrease because of radial drift.

\section{Effect of settling on the size-distribution}\label{sec:nama_diff}
\begin{figure}[h]
\centering
\includegraphics[clip=,width=.5\linewidth]{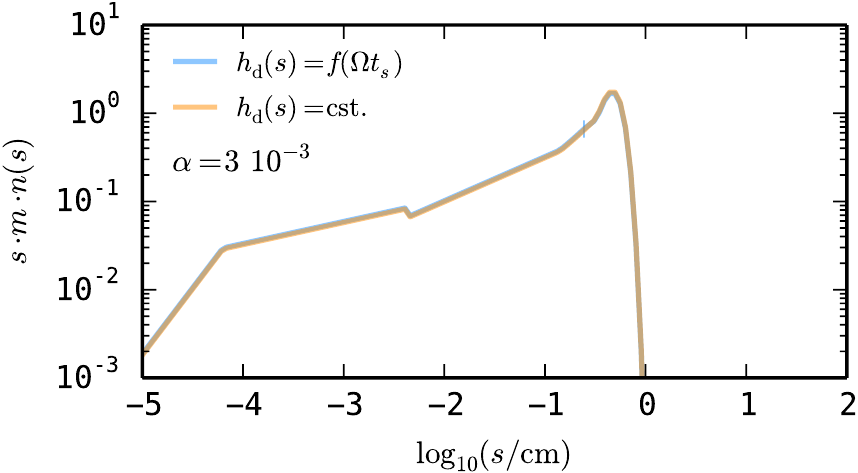}\vspace{3mm}
\includegraphics[clip=,width=.5\linewidth]{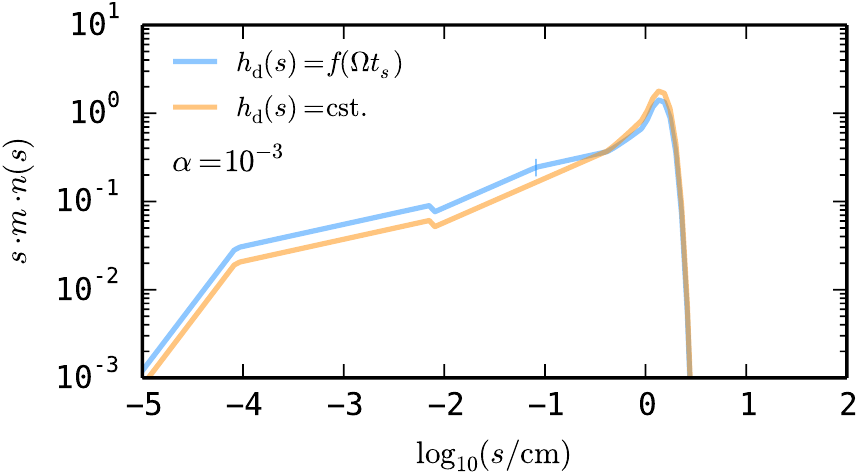}\vspace{3mm}
\includegraphics[clip=,width=.5\linewidth]{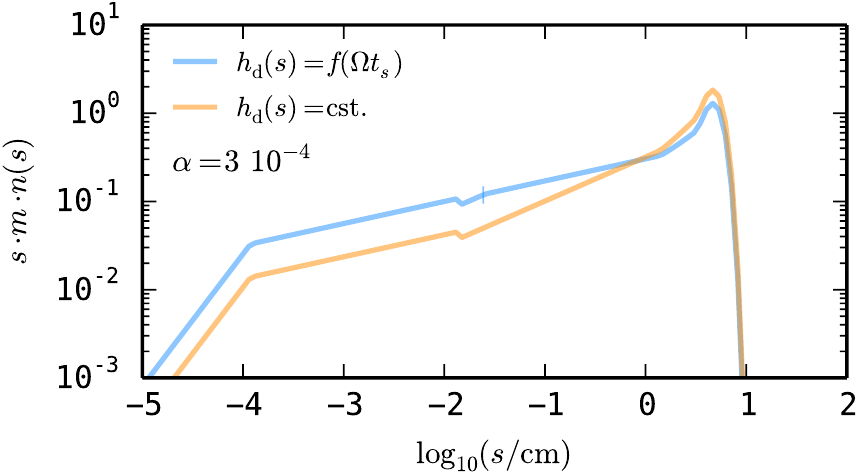}
\caption{Vertically-integrated size distributions predicted by the recipe of \citet[][Sect.~5.2]{birnstiel2011} for 3 different turbulence levels. The blue curves show the distribution when the settling is described by Eq. \ref{eq:h_d} while the orange curves correspond to the case where all dust particles share the same scale height. The settling size $s_\mathrm{sett}$ is indicated by the vertical marker. Differences are most pronounced when the turbulence is weak.}
\label{fig:nama_diff}
\end{figure}

The function of \citet{birnstiel2011} uses a variety of power-laws to build up the steady-state size distribution between a handful of important characteristic sizes (some of them are shown in Fig. \ref{fig:nama}). The dictated power-law exponents depend on whether (differential) settling is acting in a specific size regime \citep[][Table 2]{birnstiel2011}. The blue curves in Fig. \ref{fig:nama_diff} show the resulting size-distributions for different values of $\alpha$ when settling is included for particles larger than $s_\mathrm{sett}$ (indicated by the vertical marker). To mimic collisional trapping of small grains, we can set $s_\mathrm{sett}\rightarrow \infty$, effectively forcing all dust grains (small and large) to have the same scale-height. It does not matter whether this scale-height is $h_\mathrm{g}$ (as normally the case in the non-settled case) or smaller (in the case of $\Phi \ll 1$). This allows us to get a feeling for how the size-distribution itself might change when small grains are not capable of leaving the midplane. The orange curves in Fig. \ref{fig:nama_diff} show the size-distributions for equally settled dust grains. 

\end{document}